\documentclass[%
reprint,
superscriptaddress,
amsmath,amssymb,aps,pra,
floatfix,
showkeys,
]{revtex4-2}

\usepackage{graphicx}
\usepackage{dcolumn}
\usepackage{bm}
\usepackage{hyperref}
\DeclareUnicodeCharacter{00B0}{\degree}

\usepackage{xcolor}
\usepackage[normalem]{ulem} 

\begin{document}

\preprint{APS/123-QED}

\title{Optically pumped vector magnetometer using a strong bias magnetic field}

\author{Thomas Schönau}
\email{thomas.schoenau@leibniz-ipht.de}
\affiliation{Leibniz Institute of Photonic Technology, Albert-Einstein-Strasse 9, D-07745 Jena, Germany}

\author{Theo Scholtes}
\email{theo.scholtes@leibniz-ipht.de}
\affiliation{Leibniz Institute of Photonic Technology, Albert-Einstein-Strasse 9, D-07745 Jena, Germany}

\author{Florian Wittkämper}
\affiliation{Leibniz Institute of Photonic Technology, Albert-Einstein-Strasse 9, D-07745 Jena, Germany}

\author{Alexander Sekels}
\affiliation{Sekels GmbH, Dieselstraße 6, D-61239 Ober-Mörlen, Germany}

\author{Stefan Hiebel}
\affiliation{Sekels GmbH, Dieselstraße 6, D-61239 Ober-Mörlen, Germany}

\author{Gregor Oelsner}
\affiliation{Leibniz Institute of Photonic Technology, Albert-Einstein-Strasse 9, D-07745 Jena, Germany}

\author{Ronny Stolz}
\affiliation{Leibniz Institute of Photonic Technology, Albert-Einstein-Strasse 9, D-07745 Jena, Germany}

\date{\today}

\begin{abstract}
We present a novel approach allowing an optically pumped magnetometer (OPM) to be operated within Earth's magnetic field as a vector magnetometer whose sensitive axis can be freely defined.
This approach enables the measurement of any vector component of the Earth's magnetic field with the same sensitivity.
The OPM is realized by a microfabricated cesium vapor cell with nitrogen buffer gas, which is immersed into a homogeneous bias field of about $730\,\mathrm{\mu T}$.
Since this bias field is about one order of magnitude stronger than Earth’s magnetic field, it defines the sensitive axis of the OPM.
The bias field is generated by solid-state magnets and was designed to exhibit a very low relative inhomogeneity ($<10^{-4}$ in relative units) within the vapor cell dimensions as well as a point of vanishing temperature dependence at around $\mathrm{40^{\circ}C}$.
The OPM utilizes the light narrowing effect, which enables effective suppression of spin-exchange relaxation even in such large magnetic field amplitude.
Based on this implementation, we demonstrate a white noise floor of below $60\,\mathrm{fT/\sqrt{Hz}}$ in the interval between $\mathrm{100\,Hz}$ and $\mathrm{600\,Hz}$ and a sensor bandwidth of $>\mathrm{2\,kHz}$. 
Our approach enables unshielded ultrasensitive vectorial measurement capabilities relevant in many important applications.
\end{abstract}

\maketitle

\section{\label{sec:introduction}Introduction}
Ultrasensitive magnetic field detection plays a key role in many areas of research and industrial applications. 
It is used, e.g., in medicine for the detection of biomagnetic signals of the brain \cite{Hamalainen1993} and heart \cite{Koch2004}, in non-destructive testing of material properties \cite{Hohmann1997,Koss2022}, and various geophysical applications. 
These include mineral and natural oil/gas exploration \cite{Sharma1987}, unexploded ordnance (UXO) detection \cite{Billings2004,Munschy2007}, archaeology \cite{Clark1990,Fassbinder2015}, and many other fields.

For these applications, magnetometers based on superconducting quantum interference devices (SQUIDs) have long been considered the gold standard due to their exceptional sensitivity and large bandwidth. 
For example, a white magnetic noise of $\mathrm{0.09\,fT/\sqrt{Hz}}$ was demonstrated using a $\mathrm{29.5\,mm \times 33.5\,mm}$ sized pickup loop \cite{Schmelz2016}.
However, in recent decades, optically pumped magnetometers (OPMs) have undergone significant advancements and can now rival SQUIDs in terms of noise performance. 
These improvements primarily involve the suppression of spin-exchange relaxation between alkali metal atoms, which substantially reduces the transverse spin relaxation rate $\Gamma_2$ and consequently enhances the magnetic field sensitivity of OPMs \cite{Allred2002}. 
The suppression of spin-exchange can be achieved by two established modes of operation: The first is the spin exchange relaxation-free (SERF) mode \cite{Allred2002,Kominis2003} which requires the SERF-OPM to be operated in a near-zero magnetic field of typically less than a few nanoteslas. 
Thus, SERF-OPMs must be operated within shielded environment or by applying active magnetic field compensation \cite{Seltzer2004,Bertrand2021}. 
To our knowledge, the best noise performance of a SERF-OPM demonstrated is $\mathrm{0.16\,fT/\sqrt{Hz}}$ \cite{Dang2010}, using a measurement volume of $\mathrm{0.45\,cm^{3}}$. 
An alternative way of suppressing spin exchange relaxation is the light narrowing (LN) regime, which requires to pump most of the alkali atoms into a stretched state \cite{Appelt1999}, thus requiring a high pump laser power. 
The ability to operate in geomagnetic field strengths and above is the main advantage of LN-OPMs which are hence suited for geomagnetic applications. 
The potential of the LN phenomenon becomes apparent when gradiometric setups \cite{Smullin2009,Sheng2013,Limes2020} are considered, in which the technical noise can be strongly suppressed. 
In these cases, excellently low intrinsic noise could be demonstrated which even reached $\mathrm{0.54\,fT/\sqrt{Hz}}$ \cite{Sheng2013}. 
In a more realistic scenario involving a portable magnetometer operated in a magnetic field of $\mathrm{50\,\mu T}$, $\mathrm{140\,fT/\sqrt{Hz}}$ were achieved, with technical noise representing the limiting factor \cite{Oelsner2022}.

When magnetic sources are sought to be detected and localized, SQUIDs have the advantage of measuring the vector component perpendicular to their effective pickup area. 
With the additional information content, the solution space in inversion calculations and magnetic interpretations is reduced and hence the numerical effort is much smaller compared to using scalar OPM data. 
As OPMs commonly used in the field are scalar sensors, they lack sensitivity to field fluctuations perpendicular to $\vec{B}$ as the sensitive axis of the scalar magnetometer is usually determined by Earth’s magnetic field vector and cannot be controlled by the operator. 
This fact is illustrated by decomposing a magnetic fluctuation $\Delta \vec{B} = \Delta \vec{B}_{\parallel}+\Delta\vec{B}_{\perp}$ into the components parallel and perpendicular to $\vec{B}$, respectively:
\begin{equation}\label{eq:mf}
|\vec{B}+\Delta \vec{B} | \approx | \vec{B}+ \Delta \vec{B}_{\parallel}| + \frac{|\vec{B}_{\perp}|^2}{2 | \vec{B} |}\,.
\end{equation}
For example, a variation of $\mathrm{1\,nT}$ perpendicular to the Earth’s magnetic field (assuming $\mathrm{50\,\mu T}$) results in a measurable change of only $\mathrm{10\,fT}$ for a scalar magnetometer. 
Thus, in typical scenarios where magnetic field fluctuations are significantly weaker than the Earth's magnetic field, scalar magnetometers effectively measure only the projection of these fluctuations onto the Earth's magnetic field vector. 
To be able to adequately replace SQUIDs in geomagnetic applications such as transient electromagnetics (TEM) \cite{Nabighian1991}, OPMs thus require to measure Earth's magnetic field components along any spatial direction with a sensitivity significantly below $\mathrm{1\,pT/\sqrt{Hz}}$, as otherwise fluxgate magnetometers could be used instead. 
At the same time, they must cover a bandwidth from DC to at least $\mathrm{1\,kHz}$ to be suitable also for electromagnetic methods in geophysics.

Several approaches have been documented in the literature which in principle enable vectorial measurements using OPMs operated in finite magnetic fields: One method to obtain full vector information involves monitoring several projections of the spin dynamics using probe lasers oriented in different directions \cite{Fairweather1972,Sun2017,Afach2015}. 
Further approaches are relating the magnetic field orientation to modulation amplitudes in the probe beam, caused by magnetically sensitive resonances in electromagnetically induced transparency (EIT) \cite{Lee1998,Cox2011}, resonances in nonlinear magneto-optical rotation with frequency- or amplitude-modulated light \cite{Pustelny2006,Acosta2006}, free alignment precession \cite{Lenci2014}, modulation of the magnetic field by coils \cite{Seltzer2004,Alexandrov2004,Vershovskii2004}, parametric modulation of the longitudinal magnetic field \cite{Zhang2023}, modulation of the magnetic field by light shifts introduced by lasers from different directions \cite{Patton2014}, and parametric resonances of $\mathrm{^{4}He}$ metastable atoms in an actively compensated near-zero magnetic field \cite{Bertrand2021}.

Except \cite{Bertrand2021}, all of above mentioned approaches fail to achieve the required isotropic sensitivity. 
The reason for this seems to lie in the high demands on the relative measurement accuracy, which cannot be satisfied by amplitude measurements. 
According to \cite{Bertrand2021}, a vector magnetometer that exhibits noise significantly below $\mathrm{1\,pT/\sqrt{Hz}}$ and simultaneously covers a measurement range of approximately $\mathrm{\pm70\,\mu T}$ requires a digital-to-analog converter with huge vertical resolution and a very low noise current source for generation of the compensation field.
To circumvent these dynamic range demands, in this work we use an approach where each vector component is separately reduced to a high-precision frequency measurement \cite{PatentOPTEM}. 
This is achieved by exploiting the projection behavior described by Eq.~(\ref{eq:mf}): By exposing a conventional scalar LN-OPM to a very homogeneous and temporally stable bias field, whose field strength exceeds that of the Earth's magnetic field by at least one order of magnitude, it is basically sensitive only to the vector component parallel to the bias field. 
In other words, a single-axis vector magnetometer is obtained. 
Small systematic measurement errors arising from magnetic field variations perpendicular to the sensitive axis due to the quadratic terms in Eq.~(\ref{eq:mf}) can be corrected, if necessary, by employing a set of three such single-axis vector magnetometers arranged orthogonally to each other. 

The article is structured as follows: Section \ref{sec:highfield} outlines the influence of the non-linear Zeeman effect in strong magnetic fields and presents a compensation strategy allowing for high sensitivity of the magnetometer. The experimental setup is introduced in Section \ref{sec:expsetup} and Section \ref{sec:results} discusses the experimental results. In Section \ref{sec:summary}, the results are summarized and a conclusion is drawn regarding the potential of this new vectorial measurement approach.
\section{\label{sec:highfield}OPM in high magnetic field}
From the Breit-Rabi equation \cite{Breit1931,Woodgate1980}, an approximation for the Zeeman transitions corresponding to $\Delta m_F = 1$ is obtained by a series expansion at $B = 0$ and neglecting the small contribution $g_I \mu_N \ll g_S \mu_B$:
\begin{equation}\label{eq:breit}
\Delta E_{m_F +1} - \Delta E_{m_F} = \pm \frac{g_S \mu_B}{1+2I}B \mp \frac{g_S^2\mu_B^2 (1+2 m_F)}{(1+2I)^2 h \Delta \nu} B^2\,,
\end{equation}
where $h\Delta \nu$ is the hyperfine splitting of the ground state and $I$ is the nuclear spin. 
The remaining parameters do not depend on the atom under consideration: nuclear magneton $\mu_N\approx 5.05 \cdot 10^{-27}\, \mathrm{J/T}$, the electron g-factor $g_S\approx 2$, and the Bohr magneton $\mu_B \approx 9.27 \cdot 10^{-24}\,\mathrm{J/T}$. 
As a consequence of Eq.~(\ref{eq:breit}), in weak magnetic fields, the energy splitting of the Zeeman levels can be considered linear in $B$ and independent of $m_F$, giving rise to a single transition frequency for all Zeeman coherences. 
In this case, the alkali atoms precess around the magnetic field vector $\vec{B}$ at Larmor frequency
\begin{equation}\label{eq:larmor}
f_\mathrm{L} = \frac{g_S \mu_B}{(1+2I)h}|\vec{B}|=\frac{\gamma_\mathrm{L}}{2\pi}\,|\vec{B}|\,,
\end{equation}
where $\gamma_\mathrm{L}$ is the gyromagnetic ratio ($\gamma_\mathrm{L}\approx 2 \pi \cdot 3.5\,\mathrm{Hz/nT}$ in the electronic ground state of $\mathrm{{}^{133}Cs}$). 
In the presence of strong magnetic fields, the quadratic term in Eq.~(\ref{eq:breit}) becomes significant, leading to an approximately equidistant splitting of the transition frequencies proportional to $m_F$, which is known as the nonlinear Zeeman effect (NLZE). 
In $\mathrm{{}^{133}Cs}$ ($\Delta\nu\approx 9.19\,\mathrm{GHz}$, $I=7/2$), the splitting between two adjacent Zeeman transitions is
\begin{equation}\label{eq:deltalarmor}
\Delta f_\mathrm{L} \approx 2.67 \cdot 10^{-3}\mathrm{\frac{Hz}{\mu T^2}}\cdot B^2\,.
\end{equation}
If the frequency separation induced by this effect reaches or even exceeds the linewidth of the individual $m_F$-transitions, it can be considered as an additional line broadening or even splitting mechanism, which can considerably reduce the sensitivity of OPMs. 
This reduction in sensitivity can occur even within the Earth's magnetic field when large-volume, wall-coated vacuum cells with narrow magnetic resonance linewidths of $\lesssim 10\,\mathrm{Hz}$ are used \cite{Acosta2006,Schwindt2005}.

\begin{figure}
	\centering
	\includegraphics[width=\columnwidth]{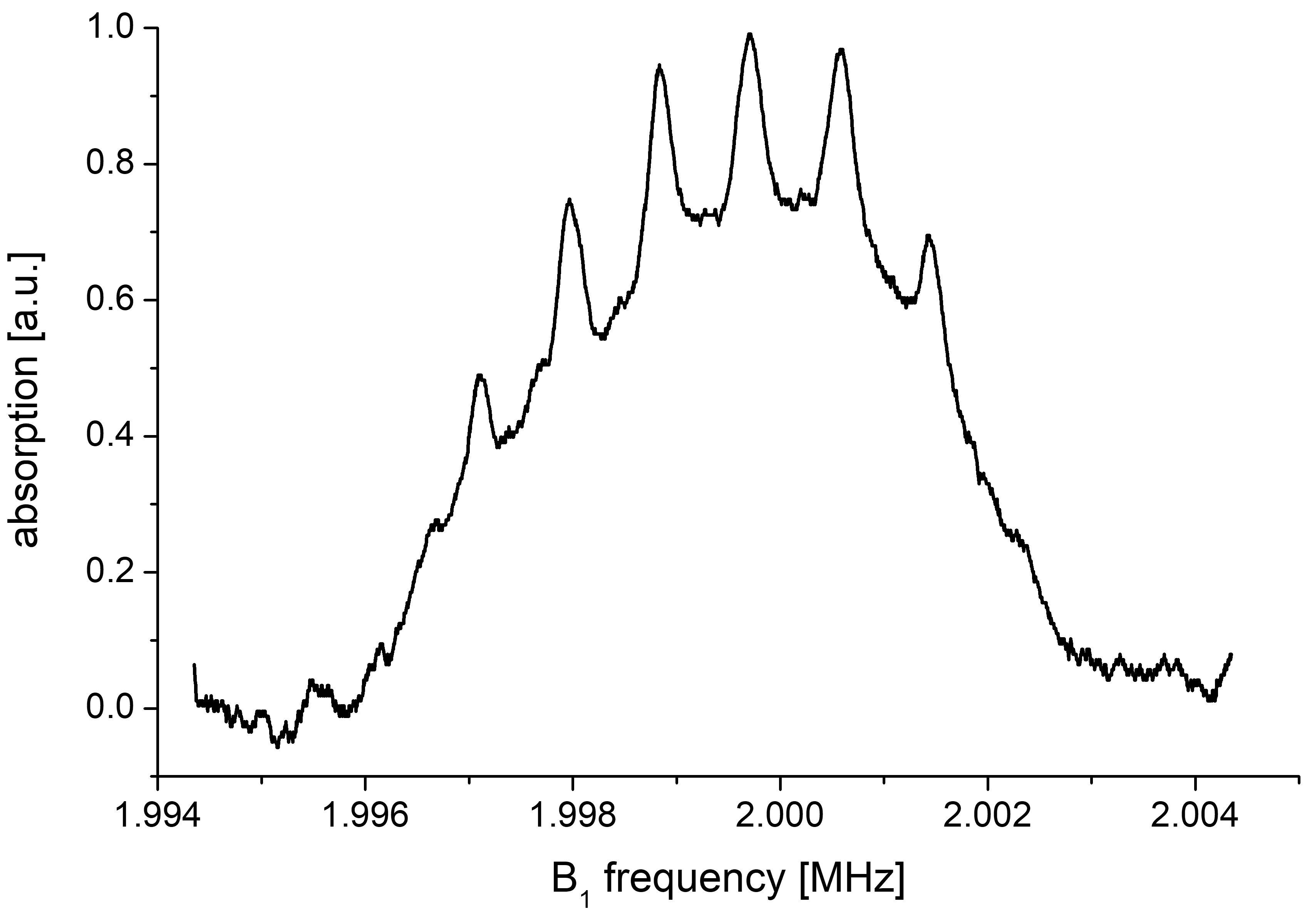}
	\caption{\label{fig:paraffin} Magnetic resonance signal of a paraffin coated Cs vacuum cell operated at room temperature, showing the nonlinear Zeeman effect at a strong magnetic field of $\approx 570\,\mathrm{\mu T}$. The laser was tuned to the $F=4 \rightarrow F'=3$ Cs D1 transition. The data was taken from \cite{Hinkel2021} for purpose of illustration. Throughout this work, a microfabricated cesium vapor cell with nitrogen buffer gas ($180\,\mathrm{mbar}$) has been used.}
\end{figure}

In case of an OPM exposed to a strong bias field $B_0>500\,\mathrm{\mu T}$, the impact of the NLZE is even more relevant, thus requiring a compensation strategy. 
For illustration purposes, the $M_z$ magnetic resonance absorption spectra on the $F=4 \rightarrow F'=3$ Cs D1 line of a paraffin-coated vacuum cell \cite{Castagna2009} operated at room temperature in a magnetic field of $570\,\mathrm{\mu T}$ are shown in Figure \ref{fig:paraffin}. 
This data was obtained during a preliminary study on the NLZE in strong magnetic fields and clearly shows the equidistant splitting of $\approx 870\,\mathrm{Hz}$ between adjacent Zeeman resonances.
Known strategies to counteract the NLZE include the application of spin-locking \cite{Bao2022}, increasing the overlap between Zeeman lines by using smaller-size buffer gas cells to increase the linewidth \cite{Schwindt2005}, or using alkali metals in which the NLZE is less pronounced. 
We use ${}^{133}\mathrm{Cs}$, which exhibits a lower NLZE in comparison to ${}^{87}\mathrm{Rb}$ or ${}^{39}\mathrm{K}$ \cite{Breit1931}, ensure sufficient overlap of Zeeman resonances by exploiting broadening mechanisms such as the use of microfabricated vapor cells, as well as focusing the magnetometer signal on a few Zeeman resonances by exploiting the LN effect in the Light-Shift-Dispersed Mz (LSD-Mz, \cite{Schultze2017,Oelsner2019a}) configuration. 

The use of relatively small size vapor cells additionally reduces the requirements on the homogeneity of the bias field $B_0$. 
Furthermore, the deployment of buffer gas instead of a spin-preserving wall coating increases both, the linewidths of the Zeeman resonances and the optical transitions.
The latter is necessary for realizing the LN effect with a single pump laser \cite{Scholtes2011}. 
By the LN effect, in turn, the $F = 3$ ground state is depleted and the population distribution is shifted towards the stretched states $|F,m_F\rangle=|4,-4\rangle$ or $|4,+4\rangle$, depending on the polarization of the pump beam ($\sigma^{-}$ or $\sigma^{+}$, respectively). 
This shift towards the stretched states significantly suppresses the spin-exchange interaction and reduces the number of Zeeman resonances that contribute to the magnetometer signal, such that useful sensitivities can be achieved despite the smaller cell volume. 

In the case considered here, the pump beam is aligned parallel to the bias magnetic field, which significantly increases the effectiveness of the pumping process. Further details on the LN effect and the LSD-Mz configuration are discussed in the next chapter and can also be found in \cite{Scholtes2016,Schultze2017,Oelsner2019a,Oelsner2022}.

Finally, by appropriately choosing the OPM operating point, it is possible to significantly broaden the Zeeman resonances with only a minor impact on the fundamental OPM sensitivity. 
This can be qualitatively explained by the fact that a simultaneous and matched increase in both the cell temperature and the laser intensity enhances both, the width and the amplitude of the measured magnetic resonance signal. 
The increase of the latter is partially compensating for the loss of sensitivity due to the increased linewidth. 
In this paper, this is referred to as "compensated broadening", in order to distinguish it from other broadening mechanisms, like for example due to magnetic field gradients. 
The compensated broadening mechanism of a temperature increase can be illustrated using the fundamental sensitivity of an OPM, which is limited by the spin projection noise \cite{Budker2013} (in units of $\mathrm{T/\sqrt{Hz}}$):
\begin{equation}\label{eq:Bsp}
B_\mathrm{sp}=\frac{1}{\gamma_\mathrm{L}}\sqrt{\frac{\Gamma_2}{nV}}\,.
\end{equation}
This sensitivity depends on the number of atoms $nV$ in the measurement volume $V$ and the transverse relaxation rate $\Gamma_2$ of the atoms in units of Hz. 
The particle density of cesium at temperature $T$ is given by
\begin{equation}\label{eq:density}
n=\frac{1}{T} \cdot 10^{27.866+A-B/T}\,,
\end{equation}
with A = 4.165 and B = 3830 \cite{Alcock1984}. 
For temperatures at and above the optimal working point, $\Gamma_2$ is dominated by spin-exchange interactions:
\begin{equation}\label{eq:gamma2}
\Gamma_2 \approx \alpha R_\mathrm{SE} = \alpha n \bar{v} \sigma \,.
\end{equation}
This rate depends on the collision rate $R_\mathrm{SE}$ between cesium atoms, with the dimensionless factor $\alpha$ accounting for the fact that not every collision results in an interaction. 
In the desired light-narrowing regime, $\alpha \ll 1$. 
The collision rate depends on the particle density $n$ as defined above, the average thermal velocity $\bar{v}$, and the interaction cross-section $\sigma$ of the cesium atoms. 
By substituting Eqs.~(\ref{eq:density}) and~(\ref{eq:gamma2}) into Eq.~(\ref{eq:Bsp}), the dependency on the cesium density $n$ is removed, yielding $B_\mathrm{sp}\propto \sqrt{\bar{v}}\propto T^{1/4}$. 

For instance, when the temperature is raised from $110\mathrm{{}^\circ C}$ to $120\mathrm{{}^\circ C}$, the linewidth broadens by a factor of approximately 1.8, while the intrinsic noise limit only increases by 0.65\%. 
In practice, the noise is typically dominated by technical factors, leading to deviating dependencies. 
However, this example clearly illustrates that the vapor temperature has a much greater influence on the linewidth than on the sensitivity. 
The possibility of increasing the linewidth by means of compensated broadening mechanisms has further advantages as it reduces the requirements for the homogeneity of the strong bias field $B_0$ and increases the bandwidth of the OPM.
\section{\label{sec:expsetup}Experimental setup}
\subsection{\label{subsec:lsdmz}LSD-Mz measurement principle}
\begin{figure}
	\centering
	\includegraphics[width=\columnwidth]{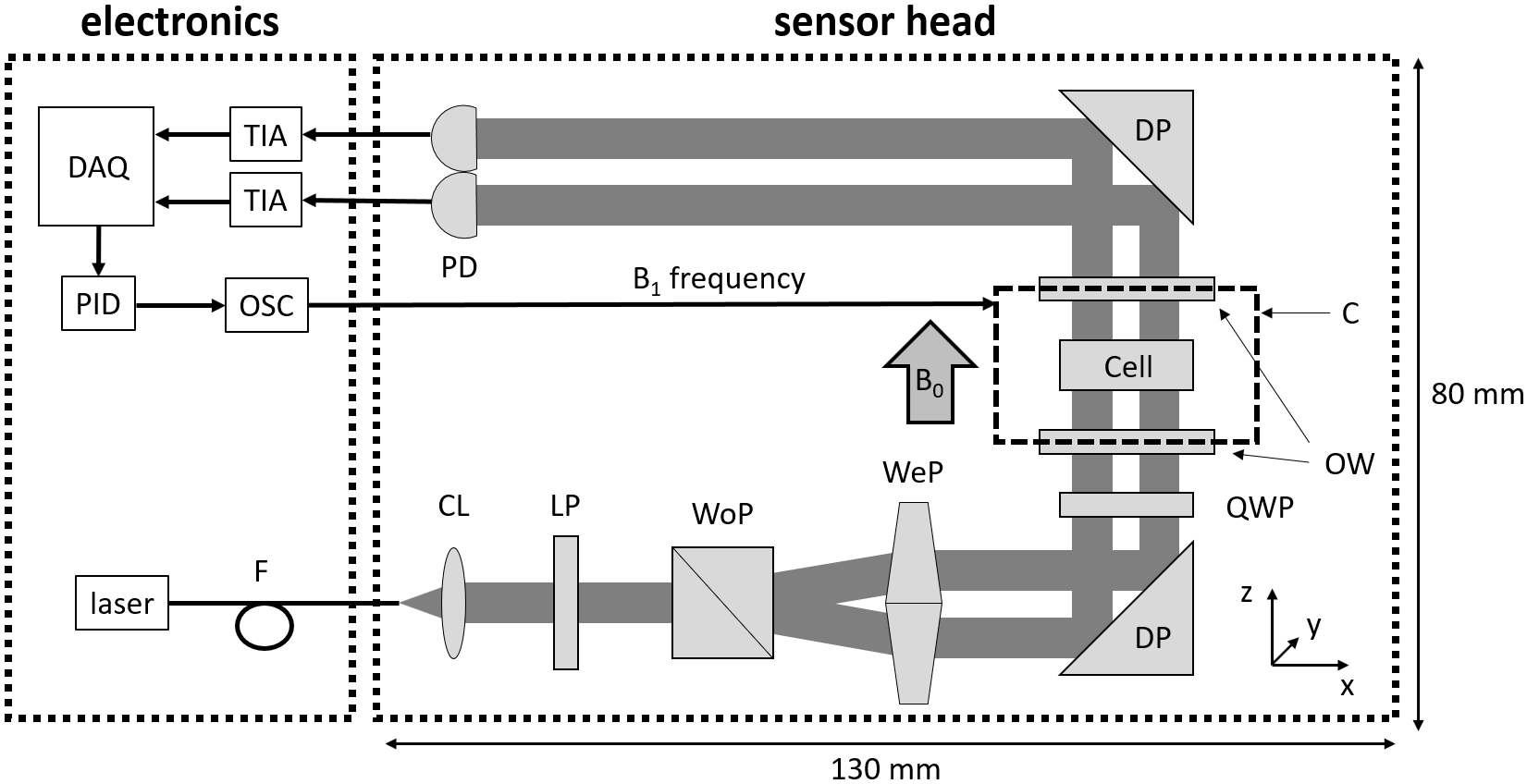}
	\caption{\label{fig:setup} Measurement setup. Details and description of the abbreviations are given in the main text.}
\end{figure}

The measurement setup is sketched in Fig.~\ref{fig:setup}, excluding the magnetic shielding barrel, the heater electronics, and the magnet system utilized for the generation of the bias magnetic field. 
The latter is implemented as a Halbach cylinder based on permanent magnets and will be described in detail in Sec.~\ref{subsec:halbach}. The sensor head is mounted on a $\mathrm{10\,mm}$ thick acrylic plate with outer dimensions of $\mathrm{130\,mm \times 80\,mm}$. 
This plate is inserted into the Halbach cylinder such that the vapor cell is placed in the central region to ensure best magnetic field homo­ge­neity.
Both, the sensor head as well as the Halbach cylinder, were installed in a three layer $\mathrm{\mu}$-metal magnetic shielding barrel of $\mathrm{1 \,m}$ in diameter and $\mathrm{1\,m}$ length, described in detail in \cite{Schultze2010}.

The laser, a Toptica DLC DL Pro designed for the Cs D1 transition at $894.6\,\mathrm{nm}$, is coupled to a polarization-maintaining single-mode optical fiber (F) by a Toptica fiber dock. 
At the sensor head, the light is collimated to a beam diameter of $\approx5\,\mathrm{mm}$ using a plano-convex lens (CL) prior to passing through a linear polarizer (LP). 
This conversion process is employed to address potential polarization variations that may arise from external factors such as vibrations or stress within the fiber, effectively transforming them into intensity variations. 
Subsequently, the horizontally and vertically polarized components of the light are split into two partial beams using a Wollaston prism (WoP). 
By rotation of the linear polarizer, the two sub-beams can be balanced for equal intensity. 
Accordingly, the plane of polarization of the optical fiber is adjusted to maximize transmission trough the linear polarizer as well. 
The two diverging sub-beams are parallelized by a pair of wedge prisms and reflected by a deflection prism (DP) in order to be aligned with the bias field $\vec{B_0}$. 
Since their polarization planes are orthogonal to each other, they can be converted into circular polarized beams with opposite helicity ($\sigma^{+}$,$\sigma^{-}$) by a properly adjusted quarter wave plate (QWP).

The vapor cell is located in a thermally isolated chamber (C) at a temperature of typically $125\mathrm{{}^\circ C}$. 
The chamber is equipped with two optical windows (OW) and a coil in a Helmholtz-like configuration, that generates the oscillating magnetic $B_1$ field in $y$-direction. 
After interaction with the Cs atoms, both beams will be reflected by a second deflection prism towards a pair of Hamamatsu S5106 photo diodes (PD). 
The measured photocurrents are converted to voltages using Femto DLPCA-200 low-noise transimpedance amplifiers (TIA) and are recorded by a data acquisition system (DAQ). 
The $B_1$ current is generated by a digital oscillator (OSC) whose frequency can be controlled by the operator or alternatively by a proportional-integral-derivative (PID) controller.

In contrast to other OPM schemes \cite{Scholtes2011,Bell1957,Bell1961}, where the detected signal is oscillating at the Larmor frequency, only a quasi-DC signal that covers the bandwidth of the OPM of a few kHz needs to be acquired. 
This significantly reduces the demands on the electronics (TIAs, DAQ). 
Especially in our case, where the Larmor frequency is around $2.6\,\mathrm{MHz}$ due to the $\mathrm{730\,\mu T}$ bias field.
Furthermore, phase-sensitive (lock-in) detection is not required, eliminating the need for potentially costly equipment and reducing the risk of systematic shifts caused by phase errors or instabilities in the electronic detection circuitry.

\begin{figure}
	\centering
	\includegraphics[width=\columnwidth]{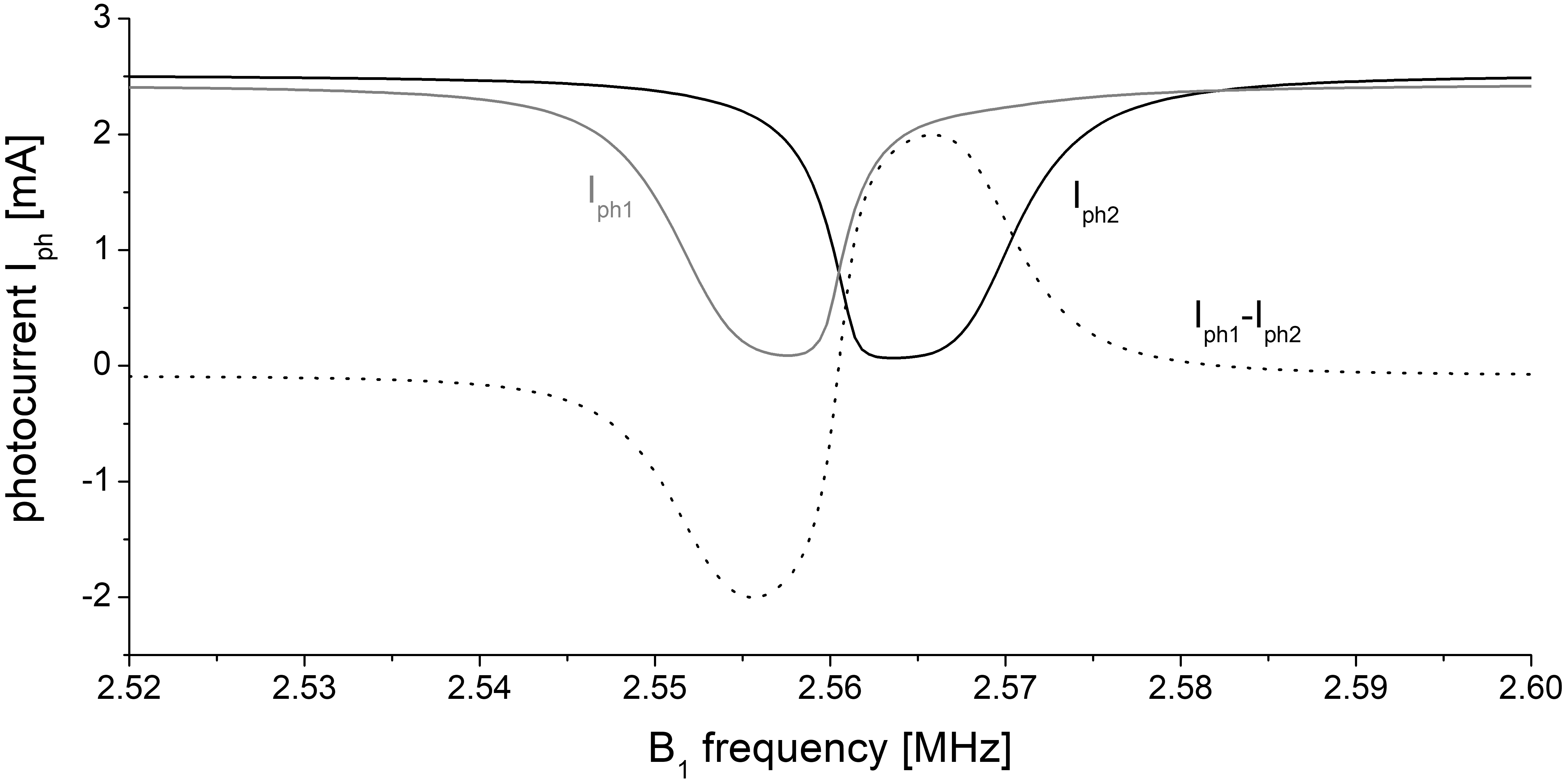}
	\caption{\label{fig:lsdmz} Magnetic resonance signal of the two beams of an LSD-Mz OPM in a bias field of $\approx 730\,\mathrm{\mu T}$ as well as the calculated difference signal.}
\end{figure}

A typical LSD-Mz magnetometer signal of our setup is depicted in Fig.~\ref{fig:lsdmz}. 
The spectra result from standard optically detected magnetic resonance (ODMR) principles and the partial beams are shifted relative to each other. 
One cause of this shift is the AC Stark effect, also known as light shift \cite{Mathur1968,Oelsner2019b} which occurs due to off-resonant pumping. 
It can be interpreted as fictitious magnetic field in the direction of the light's angular momentum.
Another cause of the observed shift is the NLZE discussed before. 
In the experimentally preferred case of equal intensities, resulting in exactly opposite fictitious magnetic fields as well as exactly opposite (${m_F} \leftrightarrow {-m_F}$) Zeeman populations, the transmission spectra are shifted symmetrically in opposite directions.
The LSD-Mz concept uses these shifts in order to obtain a dispersive-like signal by calculating the difference signal $I_\mathrm{ph1}-I_\mathrm{ph2}$ of the two beams, as indicated in Fig.~\ref{fig:lsdmz}. 
This difference signal exhibits a steep unique slope that crosses zero exactly at the Larmor frequency, assuming both beams are exactly balanced. 
The Larmor frequency can therefore be determined by operating the magnetometer in a feedback loop where a PID controller adjusts $f_{B1}$ so that $I_\mathrm{ph1}-I_\mathrm{ph2}=0$, resulting in $f_L (t) = f_{B1} (t)$.
Since only the difference signal $I_\mathrm{ph1}-I_\mathrm{ph2}$ enters into the determination of the Larmor frequency, the magnetometer is expected to be inherently robust against technical noise, like e.g. light intensity fluctuations. 
Also, systematic errors due to light shift or the NLZE which are also dependent on the pump rate, i.e. the light intensity, cancel each other out.
\subsection{\label{subsec:cell}Vapor cell}
\begin{figure}
	\centering
	\includegraphics[width=\columnwidth]{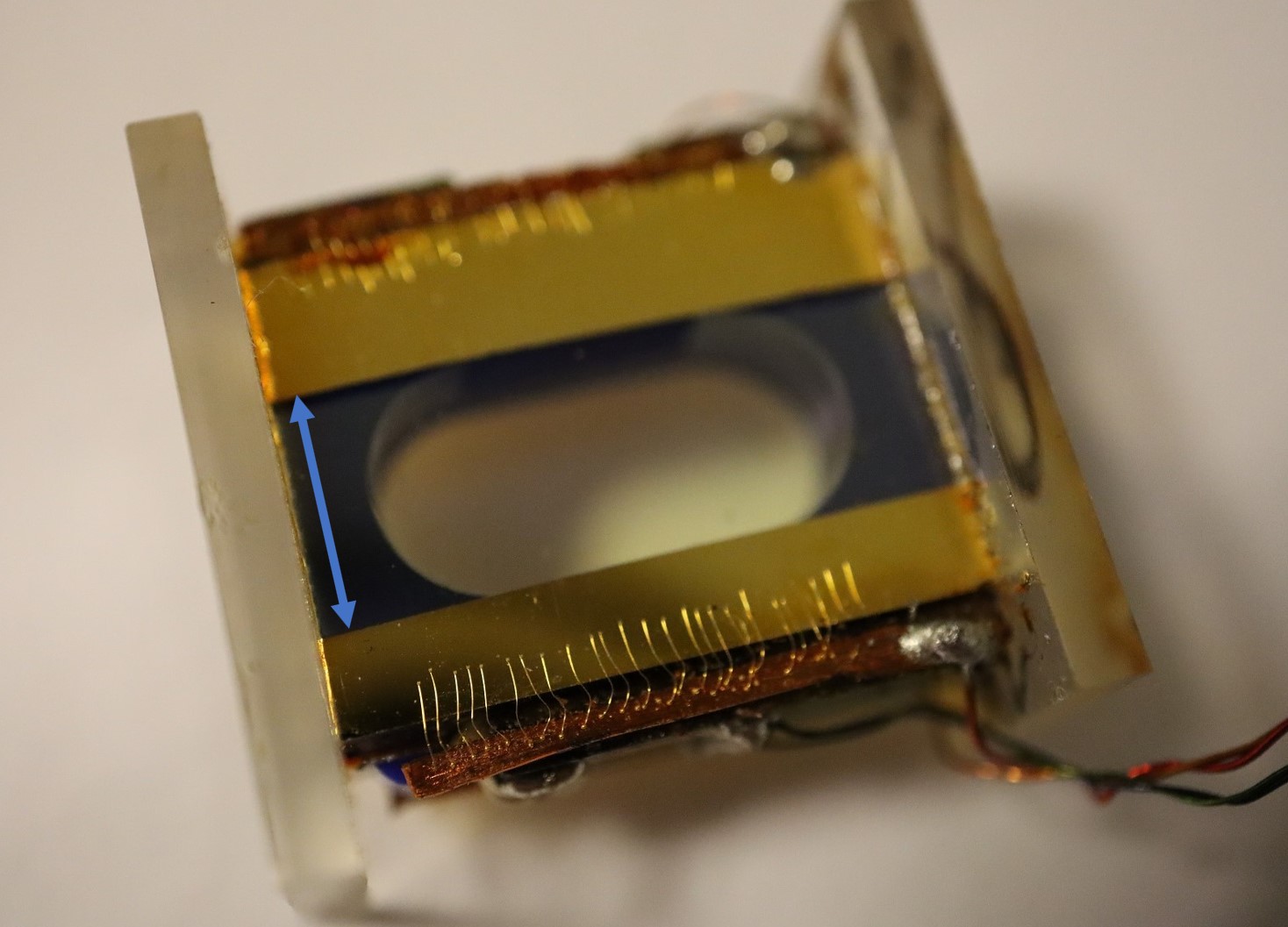}
	\caption{\label{fig:cell} Photograph of the vapor cell used in our experiments. The indicated dimension is $6\,\mathrm{mm}$. See main text for details.}
\end{figure}

The central element of the OPM sensor head is the in-house microfabricated vapor cell shown in Fig.~\ref{fig:cell}.
The cavity of this cell is specifically designed for the LSD-Mz mode to be traversed by two adjacent parallel beams. 
A detailed description of the required cell fabrication steps can be found in \cite{Woetzel2011}. 
Nitrogen with a pressure of $180\,\mathrm{mbar}$ is employed as a buffer gas, fulfilling several crucial functions.
These include reducing the collision rate of the alkali metal with the cell walls, quenching \cite{Walker1997}, and broadening the absorption profiles of the optical transitions.
A pump beam diameter of approximately $5\,\mathrm{mm}$ was chosen, resulting in a measurement volume of $157\,\mathrm{mm^3}$ regarding both beams. 
The critical volume, in which the homogeneity of the bias field $B_0$ is crucial, is defined as $|x|\leq 5.5\,\mathrm{mm}$, $|y|\leq 2.5\,\mathrm{mm}$ and $|z|\leq 2\,\mathrm{mm}$, relative to the center of the vapor cell for a beam separation of $\mathrm{1\,mm}$.

Heating resistors formed by a transparent resistive layer of indium tin oxide (ITO) are implemented directly on the outside optical windows to avoid recondensation of Cs atoms that could lead to unwanted beam scattering potentially destroying the balance of the beams. 
The dimensions of the ITO layer are approximately $\mathrm{16\,mm \times 6\,mm}$, with a thickness of $\mathrm{138\,nm}$, resulting in a resistance of approximately $\mathrm{25\,\Omega}$ for each window. 
The temperature of the vapor cell is monitored by a PT100 temperature sensor which is glued to the silicon side wall ensuring a good thermal contact. 

The alternating current used to heat the ITO layers could potentially interfere with the Larmor precession of the Cs atoms, as it may act as a secondary $B_1$ magnetic field. 
However, in our setup, this parasitic effect is significantly minimized by employing a frequency of $\mathrm{25\,MHz}$, which is well above the Larmor frequency of $\mathrm{2.56\,MHz}$. 
The heater signal is generated by a function generator, amplified by an RM Italy HLA150-Plus ham radio amplifier, and subsequently fed into the series connection of both ITO layers, which represents a well matched $\mathrm{50\,\Omega}$ load. 
In our current setup, the cell temperature is manually controlled by adjusting the output amplitude of the function generator. 
An automated PID-based temperature control loop, however, could easily be implemented in a future version. 
The high frequency of the heater does result in a non-negligible capacitive coupling to nearby electrical components, such as the photodiodes. 
However, no adverse effects were observed in the course of our experiments.
\subsection{\label{subsec:halbach}Halbach cylinder}
The measurement approach imposes specific requirements on the bias field generation to avoid compromising the performance of the OPM:
\begin{enumerate}
\item{The magnetic flux density must be in the range of $\mathrm{0.5\,mT}$ to $\mathrm{1\,mT}$, which is substantially higher than the typical strength of the Earth's magnetic field, while simultaneously avoiding unnecessarily large splitting caused by the NLZE.}
\item{The field inhomogeneity should be negligible. The specified maximum variations are $\Delta B_0 < 100\,\mathrm{nT}$ within the previously defined critical volume corresponding to a Larmor frequency variation of $350\,\mathrm{Hz}$. For comparison: according to Eq.~ (\ref{eq:deltalarmor}), the splitting in $B_0 = 500\,\mathrm{\mu T}$ is about $668\,\mathrm{Hz}$.}
\item{Temporal variations of the bias field must be negligible compared to the intrinsic noise of the OPM.}
\item{The magnetic field outside the field generating system should decay spatially as rapidly as possible to prevent the neighboring vector magnetometers from being affected by magnetic field gradients.}
\item{The dimensions of the field generation system should not exceed $\mathrm{30\,cm}$ in any spatial direction such that a future three-axis system could fit into a transport box that can be carried by two people.}
\end{enumerate}
With the above requirements, a generation of the magnetic field by coils can be ruled out according to the current state of the art.
For example, following \cite{Bertrand2021} one can expect a noise of more than $\mathrm{714\,fT/\sqrt{Hz}}$ for a bias field of $\mathrm{500\,\mu T}$. 
Therefore, solid-state magnets have to be used. 
To provide the required field strength, homogeneity, and external spatial decay for the field generation, a slightly modified version of a Halbach cylinder was designed. 
In theory, a Halbach cylinder is defined as a cylindrical shell whose remanent magnetization satisfies the angular dependence
\begin{equation}\label{eq:magnet}
\begin{split}
& B_{R,r}=B_R \cos\varphi\,,\\
& B_{R,t}=B_R \sin\varphi\,.
\end{split}
\end{equation}
Therein, $\varphi$ is the polar angle and $B_R$ the absolute value of the remanent magnetization with the radial and tangential components $B_{R,r}$ and $B_{R,t}$, respectively \cite{Bjork2011}. 
The design and implementation of such a field generation system was carried out by Sekels GmbH \cite{Sekels}.

The Halbach cylinder was approximated using six ring structures. 
Each ring structure contains eight magnets evenly spaced at angular intervals of $\Delta\varphi = 45^\circ$ along a circular radius of $\mathrm{80\,mm}$. 
The orientations of the magnets correspond to Eq.~(\ref{eq:magnet}). 
The ring structures were distributed along the cylinder axis such that the volume in which $\Delta B_0 < 100\,\mathrm{nT}$ applies is maximized. 
Including the housing, the cylinder has a diameter of $\mathrm{200\,mm}$ and a height of $\mathrm{240\,mm}$. 
Simulations confirm that the stray field outside the Halbach cylinder decays sufficiently fast to allow a multi-channel arrangement with a displacement of $\mathrm{300\,mm}$, without affecting magnetometer performance. 
In order to fulfil the homogeneity specification, a larger number of magnets were produced and characterized with regard to their magnetic moments and deviations of the angle of magnetization. 
Of those, 48 suitable ones have been selected. 
The magnets were made of a special anisotropic $\mathrm{Sm_2Co_{17}}$ alloy, showing a slightly positive temperature coefficient, unlike more common materials like neodymium or ferrite. 
This property helps to compensate for the thermal expansion of the frame structure (rings and spacers between the rings), which would otherwise weaken the bias field. 
Additionally, the frame was constructed out of ceramic material to minimize thermal expansion. 
The field generation system was designed to have a magnetic flux density of around $730\,\mathrm{\mu T}$, with temperature dependence effectively neutralized slightly above room temperature. 
\section{\label{sec:results}Experimental results}
\subsection{\label{subsec:field}Field generation}
In order to demonstrate the homogeneity of the bias magnetic field, the whole setup was placed in the three-layer magnetic shielding barrel and the sensor head was moved in $x$- and $y$-direction (see Fig. \ref{fig:setup}) relative to the Halbach cylinder while monitoring the measured Larmor frequency. 
Since the sensor head fits exactly into the holding frame of the Halbach cylinder, movement in the $z$-direction was not possible.
For the estimation of the homogeneity, it was therefore assumed that the relative homogeneities in the $y$- and $z$-directions are similar for symmetry reasons. 
By this means, we measured an average magnetic field magnitude of $731.8\,\mathrm{\mu T}$ at room temperature and could confirm that the homogeneity requirements are fulfilled. 
However, a comparison of the magnetic resonance signals shown in Fig.~\ref{fig:lsdmz}, which are significantly broader than the $350\,\mathrm{Hz}$ associated with the specified field inhomogeneity, suggests that the homogeneity requirements can be relaxed by almost one order of magnitude without compromising the sensitivity of the OPM.

\begin{figure}
	\centering
	\includegraphics[width=\columnwidth]{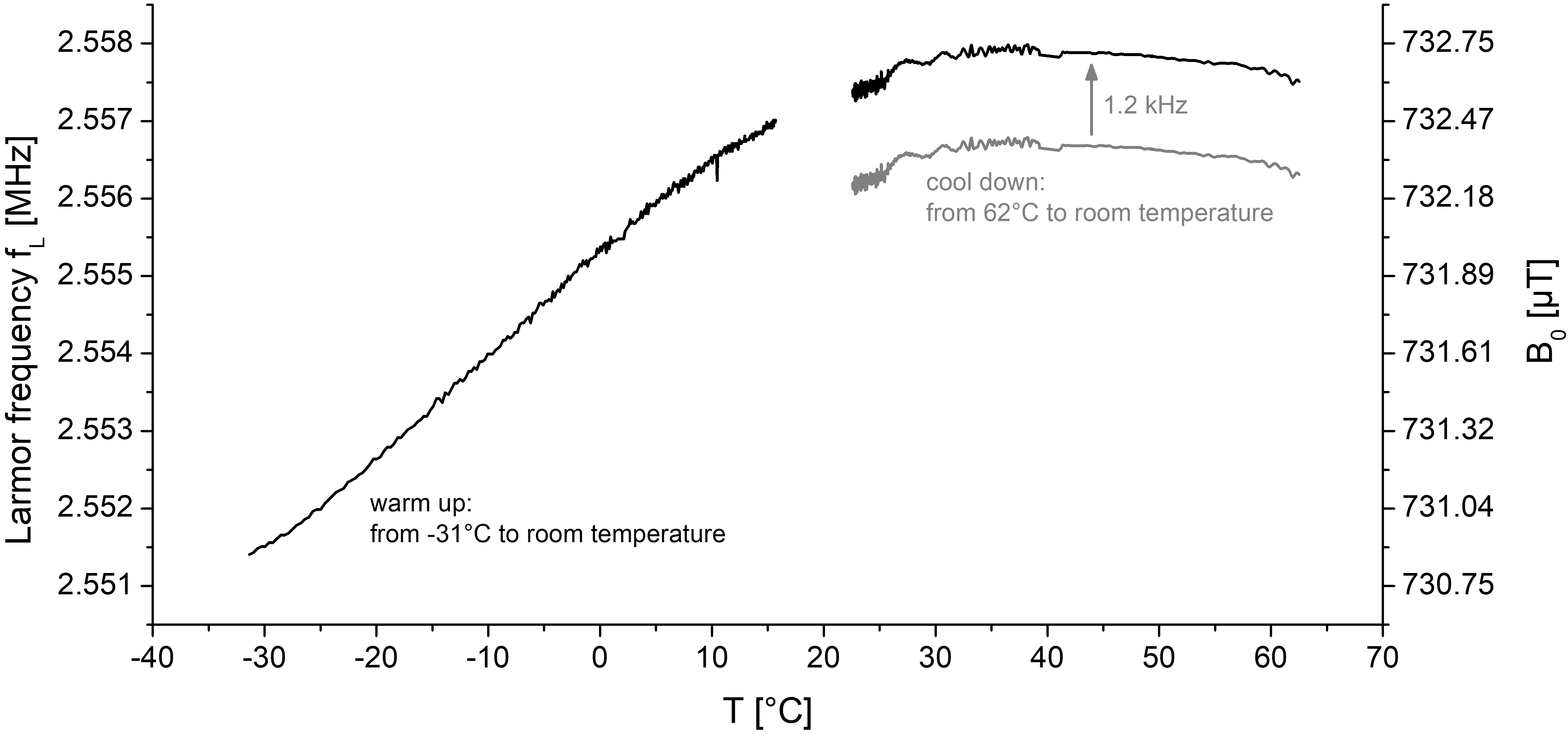}
	\caption{\label{fig:temphalbach}Temperature dependence of the magnetic bias field $B_0$. Since the warm-up and cool-down measurements were presumably superimposed by different magnetic offsets, the latter was manually shifted by $\mathrm{1.2\,kHz}$ to smooth out the dependency.}
\end{figure}

The temperature dependence of the bias field was determined by initially cooling or heating the Halbach cylinder to $-31\mathrm{{}^\circ C}$ and $+62\mathrm{{}^\circ C}$ within a climate chamber, respectively. 
After each process, the Halbach cylinder was put back into the magnetic shielding barrel within a few minutes. 
During the relaxation of the system back to room temperature, $B_0$ and the temperature were monitored using the OPM and a PT1000 temperature sensor attached to the Halbach cylinder. 
The results of the cool-down and warm-up processes are shown in Fig.~\ref{fig:temphalbach}. 
In order to minimize the time between removal from the climate chamber and the start of the measurement, the magnetic shielding barrel was left open on one side (without a lid). 
As a consequence, both measurements are slightly affected by temporal variations of the magnetic field in the laboratory as well as by different offsets due to varying positions in the open shielding barrel. 
According to Fig.~\ref{fig:temphalbach}, the temperature dependence exhibits a wide plateau region, with a maximum observed at approximately $40\mathrm{{}^\circ C}$. 
We therefore assume that measurable temperature-related fluctuations in the bias field can be minimized by maintaining the Halbach cylinder at a constant temperature close to $40\mathrm{{}^\circ C}$. 
The heating could be realized using a high-frequency current similar to that employed for cell heating, ensuring that the alternating magnetic field does not interfere with the measurements.
\subsection{\label{subsec:vectorial}Vectorial measurement principle}
\begin{figure}
	\centering
	\includegraphics[width=\columnwidth]{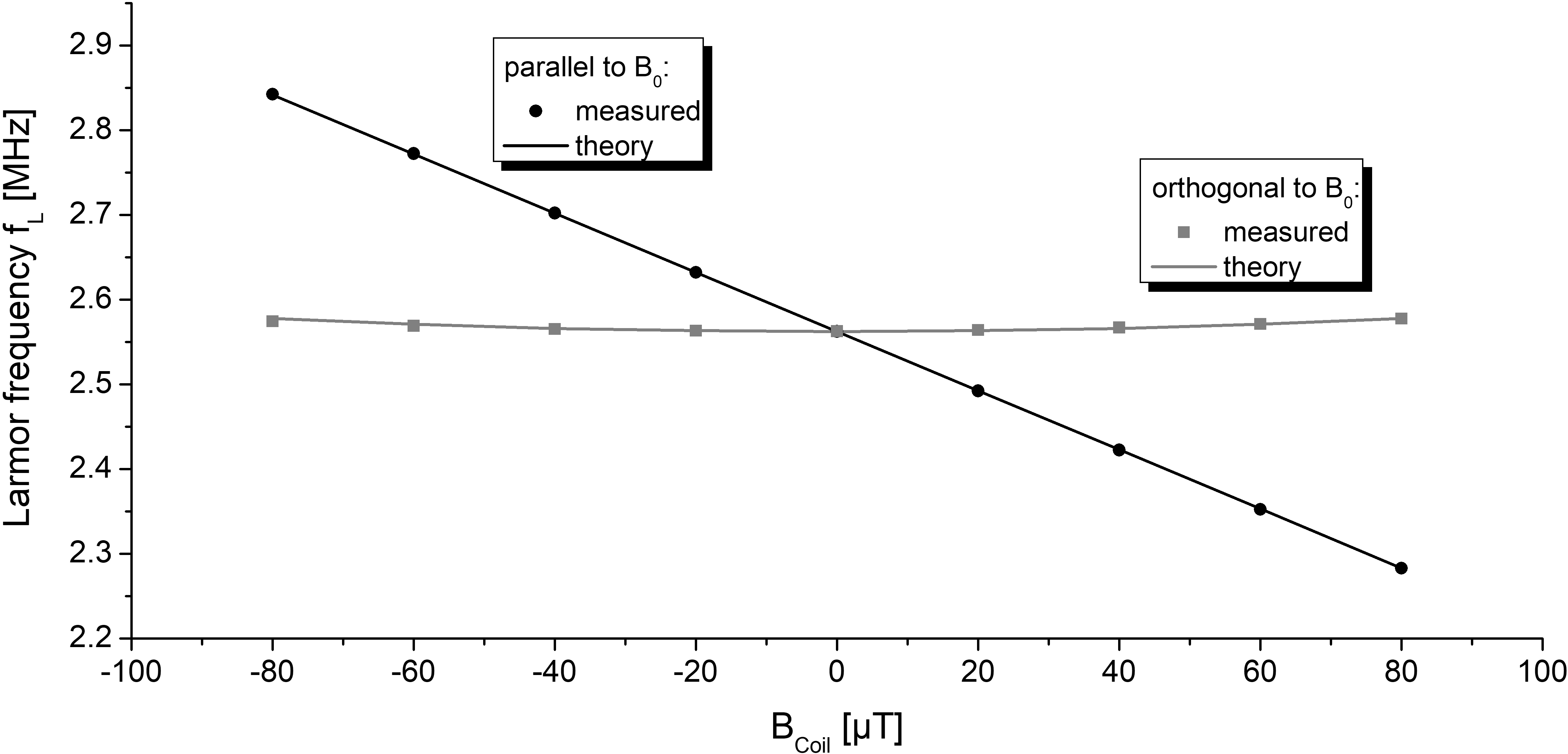}
	\caption{\label{fig:badd} Proof of the vectorial measurement principle: For a magnetic field applied parallel to the bias field $B_0$, the Larmor frequency increases linearly. Orthogonally applied magnetic fields, in contrast, are strongly suppressed, exhibiting only a very weak quadratic characteristic.}
\end{figure}

The vectorial measurement principle is verified by the data shown in Fig.~\ref{fig:badd}.
Therein, a magnetic field was applied by a set of mutually orthogonal Helmholtz coils inside the shielding barrel. 
If the magnetic field of the coils is applied in parallel to $B_0$, the measured Larmor frequency exhibits a linear dependence. 
If, in contrast, the field is applied perpendicular to $B_0$, the Larmor frequency is only slightly affected by some weak quadratic perturbation. 
The measured data fits well to the theoretical expectations represented by Eq.~(\ref{eq:mf}).
\subsection{\label{subsec:sensitivity}Sensitivity}
\begin{figure}
	\centering
	\includegraphics[width=\columnwidth]{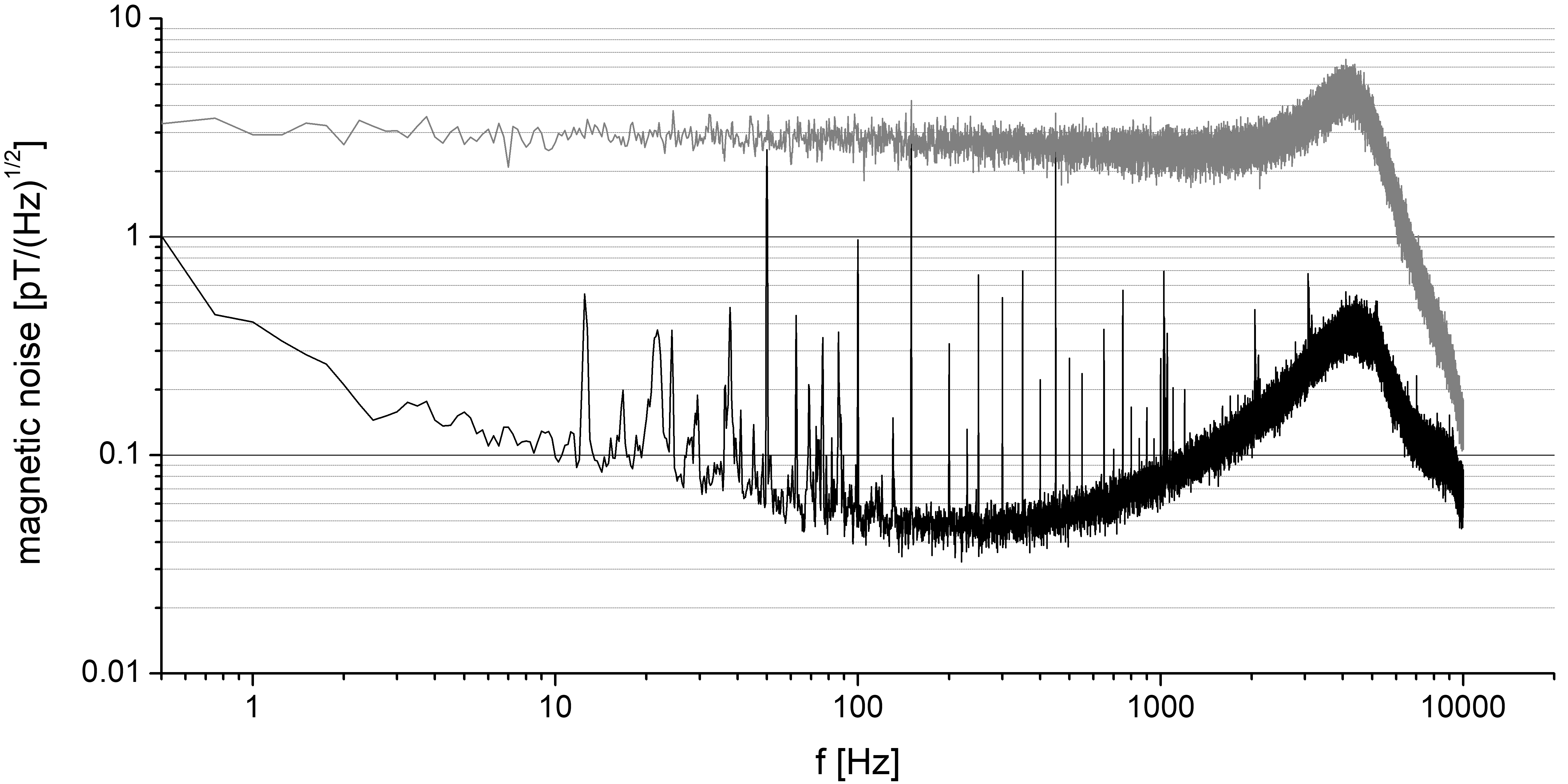}
	\caption{\label{fig:noise} Magnetic noise (black curve) of the vector magnetometer located in the three-layer shielding barrel. The OPM is exposed to a bias magnetic field of $B_0 \approx 730\,\mathrm{\mu T}$ created by the Halbach cylinder. The bandwidth of the closed-loop operated magnetometer can be determined from the grey curve, indicating the response to a white noise magnetic test signal.}
\end{figure}

In order to determine optimum sensitivity, first we recorded magnetic resonance signals as shown in Fig.~\ref{fig:lsdmz} and calculated the shot noise-limited sensitivity according to
\begin{equation}\label{eq:shotnoise}
B_\mathrm{sn} = \frac{2\pi}{\gamma_\mathrm{L}} \frac{\sqrt{2e(I_\mathrm{ph1}+ I_\mathrm{ph2})}}{s}\,,
\end{equation}
where $e$ is the elementary charge and $s=\partial(I_\mathrm{ph1}-I_\mathrm{ph2})/\partial f_{B1}$ is the slope of the differential signal at the operating point. 
In Eq.~(\ref{eq:shotnoise}), $I_\mathrm{ph1}=I_{ph2}$ can be assumed, corresponding to operation in closed feedback-loop. 
By varying laser power, $B_1$ field amplitude as well as temperature, we selected the operating point with lowest shot noise-limited sensitivity and closed the feedback-loop to directly measure the magnetic noise. 
Varying the parameters in feedback mode, we aimed for minimal measured magnetic noise and obtained a noise spectrum shown in Fig.~\ref{fig:noise}. 
Therein, the measured equivalent magnetic field noise is below $60\,\mathrm{fT/\sqrt{Hz}}$ in an interval between $100\,\mathrm{Hz}$ and $600\,\mathrm{Hz}$ and about $400\,\mathrm{fT/\sqrt{Hz}}$ at $1\,\mathrm{Hz}$. 
The operating point for this measurement was as follows: $1/e^2$ beam diameter $5\,\mathrm{mm}$, laser power at the vapor cell $\approx 9\,\mathrm{mW}$ for each sub-beam, $T = 125\mathrm{{}^\circ C}$ and $B_1 = 2.33\,\mathrm{\mu T}$ (peak). 
As we were operating at the limit of laser power, we assume that even better values could be achieved, e.g. by improving the coupling of light into the fiber. 
Between $\mathrm{10\,Hz}$ and $\mathrm{100\,Hz}$, the spectrum is dominated by mechanical vibrations of the shielding barrel, which have been already reported in SQUID measurements \cite{Schultze2010}. 
Above about $\mathrm{400\,Hz}$, the noise increases due to the limited intrinsic bandwidth of the OPM. 
However, the system bandwidth is significantly extended by operating the OPM in a feedback loop, as demonstrated in Fig.~\ref{fig:noise}. 
For the purpose of bandwidth determination, a white noise magnetic test signal was coupled into the magnetometer setup via a small coil. 
It can be seen that the response is flat in the frequency range between DC and $\approx 2\,\mathrm{kHz}$.

In order to be able to interpret the measured noise, it is compared with the theoretical values of the fundamental and the shot noise-limited sensitivity according to Eqs.~(\ref{eq:Bsp}) and (\ref{eq:shotnoise}). 
Typically, the transverse relaxation rate $\Gamma_2$ can be obtained from the linewidth of the magnetic resonance signal as shown in Fig.~\ref{fig:lsdmz}. 
In our case, however, the linewidth is considerably broadened by the NLZE and the strong $B_1$ amplitude. 
Thus, an accurate determination of $\Gamma_2$ requires more effort.
For our purposes, however, it is sufficient to estimate $\Gamma_2<10\,\mathrm{kHz}$ from Fig.~\ref{fig:lsdmz}, resulting in $B_\mathrm{sp}<1.5\,\mathrm{fT/\sqrt{Hz}}$ in the case of $T=125\mathrm{{}^\circ C}$ and $V=157\mathrm{mm^3}$. 
The shot-noise limited sensitivity can be calculated using the values $I_\mathrm{ph1}=I_\mathrm{ph2}\approx 0.83\,\mathrm{mA}$ and $s\approx 1.3\,\mathrm{\mu A/Hz}$ from Fig.~\ref{fig:lsdmz}, resulting in $B_\mathrm{sn}\approx 5.1\,\mathrm{fT/\sqrt{Hz}}$. 
These values indicate that the measured magnetic noise is at least one order of magnitude above the fundamental limits. 
The most obvious explanation is insufficient balancing of the two sub-beams. 
In this case, for example, technical noise originating from the laser is not completely compensated. 
An important task for future work in this field is therefore to improve the balancing of the beams and to identify and eliminate further potential noise sources. 
\section{\label{sec:summary}Summary}
A new approach for an OPM-based vector magnetometer was presented, in which the vector component to be measured is traced back to a high-precision frequency measurement and therefore offers unique potential in terms of dynamic range and sensitivity.
The vector magnetometer is designed to operate in Earth’s magnetic field with the intention to replace SQUID-based magnetometers in several geomagnetic applications. 
The vectorial measurement is achieved by exposing the OPM to a strong bias field of $ 732\,\mathrm{\mu T}$, which defines the sensitive axis of the vector magnetometer and furthermore prevents the occurrence of dead zones with reduced magnetometer sensitivity. 
In this bias field, a magnetic noise of $< 60\,\mathrm{fT/\sqrt{Hz}}$ was achieved in the interval between $\mathrm{100\,Hz}$ and $\mathrm{600\,Hz}$, which to our knowledge has never been reached before in terms of relative sensitivity. 
Since the estimated shot noise-limited sensitivity of $5.1\,\mathrm{fT/\sqrt{Hz}}$ is still about one order of magnitude below the measured noise, the latter seems to be dominated by technical noise, thus offering a huge potential of further sensitivity improvements.  
In closed-loop operation, the vector magnetometer offers a bandwidth of $>2\,\mathrm{kHz}$. 
This unique combination of vectorial sensitivity, low noise, high bandwidth and easier handling makes our approach a promising alternative to SQUID-based magnetometers for various applications.

In this work, we have chosen the LSD-Mz configuration as it naturally fits to the requirement of working in a strong magnetic field of well-defined direction. 
Moreover, we consider it to be best suited for dealing with the consequences arising from the nonlinear Zeeman splitting. 
However, the basic idea of achieving a vector magnetometer by applying a strong bias magnetic field to the OPM might also work for other operational schemes.

An excellent homogeneity and temporal stability of the bias field are essential for the performance of the vector magnetometer. 
For this reason, the bias field was created by a Halbach cylinder based on precision solid-state magnets with a defined temperature coefficient. 
This configuration enabled an excellent homogeneity of $\Delta B_0<100\,\mathrm{nT}$ within the region of interest and a vanishing temperature dependence $B_0 (T)$ without adding excess magnetic field noise. 

In future work, we will analyze a further simplification of the Halbach cylinder and characterize the long-term stability of the bias field as it determines the range of application for which this type of sensor is suitable. 
For our actually intended geomagnetic application scenario, the transient electromagnetic method (TEM), a reproducibility of a few minutes is sufficient and has already been achieved. 
Currently, we are working on portable electronics to replace the bulky laboratory equipment and allow for operation outside of the lab.

%
\begin{acknowledgments}
We gratefully acknowledge financial support from the Federal Ministry of Education and Research (BMBF) of Germany under Grant No. 13N15436 (OPTEM).
\end{acknowledgments}
\bibliographystyle{apsrev4-2}
%
%
%
%
\end{document}